# ACROBAT - a multi-stain breast cancer histological whole-slide-image data set from routine diagnostics for computational pathology

## Authors

Philippe Weitz[*], Masi Valkonen[*], Leslie Solorzano[*], Circe Carr, Kimmo Kartasalo, Constance Boissin, Sonja Koivukoski, Aino Kuusela, Dusan Rasic, Yanbo Feng, Sandra Kristiane Sinius Pouplier, Abhinav Sharma, Kajsa Ledesma Eriksson, Leena Latonen, Anne-Vibeke Laenkholm, Johan Hartman[§], Pekka Ruusuvuori[§], Mattias Rantalainen[§]

* contributed equally
§ contributed equally

## Abstract

The analysis of FFPE tissue sections stained with haematoxylin and eosin (H&E) or immunohistochemistry (IHC) is an essential part of the pathologic assessment of surgically resected breast cancer specimens. IHC staining has been broadly adopted into diagnostic guidelines and routine workflows to manually assess status and scoring of several established biomarkers, including ER, PGR, HER2 and KI67. However, this is a task that can also be facilitated by computational pathology image analysis methods. The research in computational pathology has recently made numerous substantial advances, often based on publicly available whole slide image (WSI) data sets. However, the field is still considerably limited by the sparsity of public data sets. In particular, there are no large, high quality publicly available data sets with WSIs of matching IHC and H&E-stained tissue sections. Here, we publish the currently largest publicly available data set of WSIs of tissue sections from surgical resection specimens from female primary breast cancer patients with matched WSIs of corresponding H&E and IHC-stained tissue, consisting of 4,212 WSIs from 1,153 patients. The primary purpose of the data set was to facilitate the ACROBAT WSI registration challenge, aiming at accurately aligning H&E and IHC images. For research in the area of image registration, automatic quantitative feedback on registration algorithm performance remains available through the ACROBAT challenge website, based on more than 37,000 manually annotated landmark pairs from 13 annotators. Beyond registration, this data set has the potential to enable many different avenues of computational pathology research, including stain-guided learning, virtual staining, unsupervised pre-training, artefact detection and stain-independent models.

## Background & Summary

Breast cancer is the most common cancer in women globally with 11.7% of all cases and the fourth most common cause of cancer deaths in women with 6.9% of all cancer deaths[1]. Biomarker assessment through IHC staining, particularly of the hormone receptors ER and PGR, for oestrogen and progesterone, respectively, as well as for the receptor for human epidermal growth factor 2, HER2, has become an essential component of the routine pathology workflow where available[2,3]. Another biomarker that is routinely assessed through IHC staining in some countries is KI67. The *International KI67 in Breast Cancer Working Group* (IKWG) currently recommends KI67 scoring at least in patients classified as ER-positive and HER2-negative based on IHC scores[4].

Automated IHC biomarker scoring with image analysis software can enhance its validity and reproducibility. The IKWG found that e.g. for KI67 scoring, automated scoring with QuPath[5] shows outstanding reproducibility[6]. In recent years, image analysis in the context of computational pathology has advanced for a whole range of applications. This has at least in part been facilitated by large, publicly available WSI data sets, such as the resources provided by the TCGA research network. Publicly available data does not only provide development data to the research community, but perhaps even more importantly, it also allows for a comparable benchmarking of novel methods on the same test data.

There are several application areas where multi-modal (e.g. multiple stains) WSI image data is required. This includes development of high performing WSI registration (the spatial alignment of corresponding tissue in two or more WSIs) methods, which is an enabling technology both for research and diagnostics. It can allow clinicians to fuse information from different IHC-stains in WSI viewers. Combining information from H&E-stained tissue with corresponding IHC-stained tissue regions can e.g. be of critical importance when investigating resection borders with respect to malignancy, which can be very time consuming without tissue alignment. Some commercial IHC scoring softwares also align H&E and IHC-stained tissue to enable pathologists to contextualise automated scoring results. In research projects, WSI registration can facilitate stain-guided learning[7–9], virtual staining[10–12], 3D reconstruction[13,14] and the transfer of annotations between different WSIs and stains. However, there is currently a lack of publicly available data sets that include WSIs from H&E-stained tissue sections with matched IHC-stained tissue from the same tumour, despite the importance of IHC for pathological diagnosis.

To promote and enable further research in this domain, we have published the ACROBAT (AutomatiC Registration Of Breast cAncer Tissue) data set[15], which consists of 4,212 WSIs from 1,153 female primary breast cancer patients. For each patient, the data set contains one WSI of H&E stained tissue and up to four WSIs with tissue that was stained with the routine diagnostic IHC markers ER, PGR, HER2 and KI67. An example of a case from the data set with all four IHC antibodies available is depicted in Fig. 1.

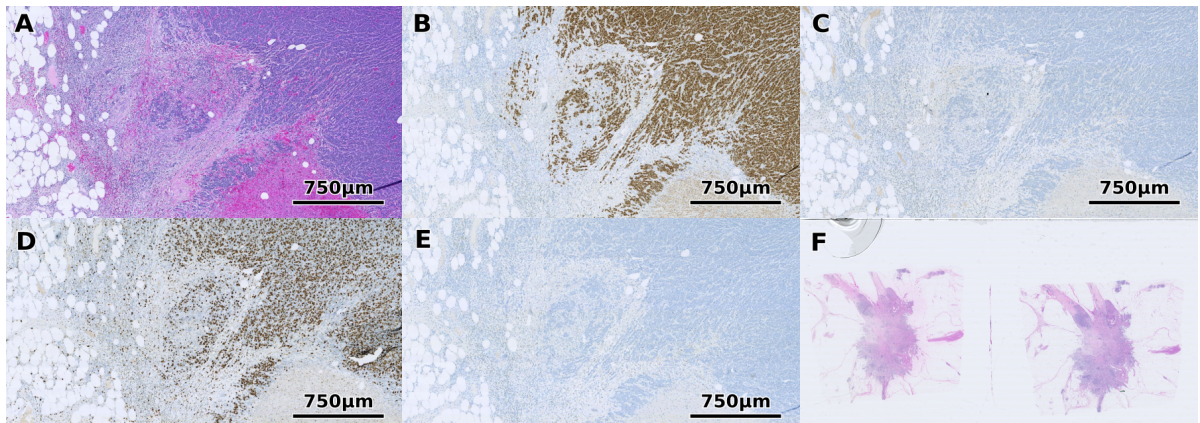

Fig. 1 *Example of an H&E-stained tissue region with corresponding IHC-stained tissue with all four routine diagnostic stains available in this data set. A) shows H&E, B) ER, C) HER2, D) Ki67 and E) PGR. F) shows an example of a WSI that was excluded since it contains multiple tissue sections.*

The data set was initially collected as part of the CHIME study (chimestudy.se) at Karolinska Institutet (Stockholm, Sweden). The primary purpose of the CHIME study is to advance precision medicine through computational pathology, based on population representative patient cohorts. Histopathology slides came from the routine clinical diagnostic workflow, with WSIs generated using high-throughput histopathology slide scanners at Karolinska Institutet.

The primary purpose of publishing this data set was to enable the ACROBAT WSI registration competition (acrobat.grand-challenge.org), which took place in the scope of the MICCAI (Medical Image Computing and Computer Assisted Intervention) 2022 conference.

While the primary purpose of this data set is the development of WSI registration methods, we believe that there could also be other use cases for the data. These may include the development of digital staining and stain transfer methods, as well as novel methods in stain-guided learning. Further applications may include the development of tissue segmentation and classification algorithms, the development of artefact detection or correction methods and unsupervised pre-training e.g. of convolutional neural networks (CNNs), which are then fine-tuned for specific tasks. We therefore hope that this data set can contribute towards the advance of WSI registration methods but also other research activities in the domain of computational pathology.

## Methods

### Data selection and splits for ACROBAT data set

The CHIME breast cancer study is based upon a retrospective cohort study design. Consecutive female breast cancer cases diagnosed between 2012 and 2018 at Södersjukhuset (Stockholm, Sweden) were included. The archived histopathology slides were retrieved and scanned. The training and validation set are a randomly selected subset of the data generated in terms of the CHIME study. The test set WSIs were chosen as a subset of the CHIME study data that has previously been reviewed by a pathologist specialising in breast pathology in the context of another research project. Cases were excluded for the ACROBAT data set only if one of the WSIs contains multiple sections of the same tissue, which occurs in approximately 1% of H&E WSIs in this data set, as depicted in Figure 1 F. These cases were excluded since it would be unclear to which of the multiple sections tissue from the corresponding IHC WSI should be aligned to. In this case, a new case was randomly selected and included after the corresponding quality control for multiple tissue sections.

|  | Training | Validation | Test | Total |
|---|---|---|---|---|
| Cases | 750 (65.0%) | 100 ( 8.7%) | 303 (26.3%) | 1153 |
| WSIs | 3406 (80.9%) | 200 ( 4.7%) | 606 (14.8%) | 4212 |
|  |  |  |  |  |
| **Stain and antibodies** |  |  |  |  |
| H&E | 750 (22.0%) | 100 (50.0%) | 303 (50.0%) | 1153 (27.4%) |
| ER | 732 (21.5%) | 29 (14.5%) | 84 (13.9%) | 845 (20.1%) |
| KI67 | 732 (21.5%) | 29 (14.5%) | 82 (13.5%) | 843 (20.0%) |
| PGR | 728 (21.4%) | 28 (14.0%) | 81 (13.4%) | 837 (19.9%) |
| HER2 | 464 (13.6%) | 14 ( 7.0%) | 56 ( 9.2%) | 534 (12.7%) |
|  |  |  |  |  |
| **Scanners** |  |  |  |  |
| NanoZoomer S360 | 559 (16.4%) | 38 (19.0%) | 205 (33.8%) | 802 (19.0%) |
| NanoZoomer XR (1) | 884 (26.0%) | 46 (23.0%) | 203 (33.5%) | 1133 (26.9%) |
| NanoZoomer XR (2) | 1963 (57.6%) | 116 (58.0%) | 198 (32.7%) | 2277 (54.1%) |

**Table 1** *Distribution of cases and WSIs to training, validation and test set, as well as respective distributions of stains, IHC antibodies and scanner models. Three Hamamatsu NanoZoomer scanners were used for WSI digitisation, one NanoZoomer S360 and two NanoZoomer XR.*

The training set consists of 3,406 WSIs from 750 patients. Each patient has one H&E WSI and up to four associated IHC WSIs from the routine diagnostic IHC antibodies ER, PGR, HER2, KI67, as depicted in Fig. 1. The validation data set consists of 200 WSIs from 100 patients and the test set consists of 606 WSIs from 303 patients. Each case in the validation and test sets consists of one H&E WSI per case and one IHC WSI, which was randomly selected stratifying for IHC antibody. The test set was furthermore selected by stratifying for clinical covariates by balanced sampling from the three different WSI scanners. Table 1 indicates the distributions of scanners and IHC antibodies in the respective subsets.

**Whole slide image scanning**
WSIs in the CHIME study that were available at the time of data selection were generated from archived histopathology slides with three Hamamatsu WSI scanners, consisting of one NanoZoomer S360 and two NanoZoomer XRs. Slides were digitised by a trained scanning technician using an automated scanning workflow, with manual rescanning of slides where automated focusing was not successful. Slides were scanned at a resolution of approximately 0.23µm/pixel and a JPEG compression quality level of 80.

**Image processing**
The 40X NDPI WSIs were first anonymized and then converted to pyramidal TIFF files with 10X and lower resolutions. Macro images and other identifying information in the WSI metadata were removed with code available from [github.com/bgilbert/anonymize-slide](github.com/bgilbert/anonymize-slide). Then, file names were generated consisting of a random case ID, the stain or antibody of the WSI, as well as the name of the respective set out of training, validation or test. TIFF files were then extracted using the libvips[16] command *im_vips2tiff* at 10X and lower magnifications, with 7 to 9 magnification levels depending on the WSI available and a downsampling factor of 2 between these levels. This reduces the storage requirements of the data set from 10.13 TB to 482 GB, likely without impacting the performance of image registration algorithms as registration is typically performed at fairly low resolutions with diminishing to no improvements at higher resolutions [17].

**Annotation workflow**

Members of the ABCAP research consortium (abcap.org) were enrolled as annotators to generate landmarks, including 13 individuals in total. All annotators have previous experience from working directly with WSIs in a research context and have received corresponding training. Two of the annotators have pathologist training. Landmark annotations were generated using a customised version of TissUUmaps[18]. Image pairs in the validation data were annotated by one annotator, whereas each image pair in the test data was annotated by two annotators in two annotation phases. All annotations were conducted with the original NDPI files at 40X magnification. Counting landmarks from both annotation phases independently, annotators generated 35,760 landmark pairs in total.

In the first phase of the annotation process, which is the same for the validation and test data, annotators were shown one H&E and one IHC stained section side-by-side and were asked to mark 50 corresponding landmarks in both images, inserting first the IHC point and then the H&E point.

In the second phase, which was only applied to the test data, annotators were provided with modified annotation files from the first round of annotations. Landmark coordinates in the IHC image were fixed in place, while for the H&E random uniform noise of [-500, 500) pixels (±115μm) was added to both the X and Y coordinates. Annotators were then asked to move the H&E landmarks to match the corresponding ones in the IHC WSI. Annotators were chosen randomly such that phase one and phase two annotations were created by different observers for each WSI. Detailed annotation guidelines for both phases are available from github.com/rantalainenGroup/ACROBAT.

**Automated evaluation of registration performance**

The data set is split into a training, a validation and a test set. In order to evaluate the performance of registration methods, landmarks for the IHC WSIs in the validation and test data, are released publicly, whereas the target H&E landmarks can be used to quantify registration performance through an automated evaluation tool. Registered validation set landmarks can be submitted at [acrobat.grand-challenge.org](acrobat.grand-challenge.org) in order to receive performance metrics based on these landmarks. It is also possible to submit registered test set landmarks on this website. However, in order to limit the number of submissions to the test data to preserve its independence and therefore usefulness, the challenge organisers need to be contacted in order to unlock test set submissions.

## Data Records

We published 4,212 WSIs of breast cancer resection specimens stained with H&E or IHC (ER, PGR, HER2, KI67) originating from 1,153 patients on the Swedish National Data Service SND ([snd.gu.se/en](snd.gu.se/en)). All WSIs are provided as pyramidal TIFF files, starting at 10X resolution (ca. 0.92 μm/pixel) and lower resolutions. The naming convention of all WSIs follows the pattern

*caseid_stain_set.tiff* where *caseid* indicates a randomly generated case ID, *stain* either H&E or the IHC antibody used, and *set* whether the file belongs to *train*, *valid* or *test*. Furthermore, there is a CSV table that indicates the microns-per-pixel at the first level in the respective TIFF files, the stain, the IHC antibody and the data split for each file name. This table is summarised in Table 1.

## Technical Validation

All WSIs originate from slides that were used in the routine diagnostic workflow. The tissue samples have therefore each been reviewed by at least one specialty pathologist using a microscope during the initial diagnosis. The macro images of all WSIs in the data set were reviewed by at least one observer in order to exclude WSIs with multiple tissue sections of the same resection specimen and in order to confirm that H&E and IHC tissue sections show corresponding tissue. All WSIs in the validation set were reviewed by at least one and all WSIs in the test set were reviewed by at least two human annotators during the landmark generation at 40X resolution. All WSIs in the test data were furthermore reviewed by a specialty pathologist while generating annotations for a research project that is independent from the ACROBAT challenge at 40X resolution, confirming their usability. The majority of WSIs in this study have also been used in other research studies, which further supports the validity of the data set[19]. There are several studies that use WSIs that were generated using the same scanners and workflow[20–22]. Some of the WSIs included in the data set contain artefacts. These WSIs were deliberately left in the data set, in order to be able to assess the robustness of suggested registration methods.

The quality of landmarks in the test data can be assessed by computing the distances between the two human annotators. Landmarks with a distance between annotators of more than 115 μm were excluded, which was chosen as a threshold in correspondence to the noise added for the second annotation phase.

## Usage Notes

Pyramidal TIFF files are compatible with OpenSlide[23] and can e.g. be inspected with QuPath. Registration algorithms typically align WSIs iteratively starting at low resolutions. With OpenSlide, lower resolution versions of the WSI can be obtained through the different levels of the TIFF files, which makes additional computations for downsampling obsolete. ACROBAT Github repository at github.com/rantalainenGroup/ACROBAT provides code to visually inspect landmarks, either only in IHC or also paired H&E landmark after registration by the user.

## Code Availability

NDPI files were anonymized with code available from github.com/bgilbert/anonymize-slide. The libvips package for pyramidal TIFF extraction is available from github.com/libvips/libvips. The tool used to generate landmark annotations is based on TissUUmaps, which is available from tissuumaps.github.io. Code used for displaying landmarks in a surrounding tissue region, code for computing registration performance metrics, as well as the annotator protocols are available from github.com/rantalainenGroup/ACROBAT.

## Acknowledgements

We acknowledge support from Stratipath and Karolinska Institutet sponsoring the ACROBAT challenge prize; MICCAI society for hosting the ACROBAT challenge, and Nguyen Thuy Duong Tran for support with digitising histopathology slides.

We acknowledge funding from:
Vetenskapsrådet (Swedish Research Council)
Cancerfonden (Swedish Cancer Society)
ERA PerMed (ERAPERMED2019-224-ABCAP)
MedTechLabs
Swedish e-science Research Centre (SeRC)
VINNOVA
SweLife

Academy of Finland (#341967, #334782, #335976, #334774)
Cancer Foundation Finland
University of Turku Graduate School
Turku University Foundation

Oskar Huttunen Foundation
David and Astrid Hägelén Foundation

## Author information

### Authors and affiliations

**Department of Medical Epidemiology and Biostatistics, Karolinska Institutet, Stockholm, Sweden**
Philippe Weitz, Leslie Solorzano, Kimmo Kartasalo, Constance Boissin, Abhinav Sharma, Kajsa Ledesma Eriksson, Mattias Rantalainen

**Institute of Biomedicine, University of Turku, Turku, Finland**
Masi Valkonen, Circe Carr, Aino Kuusela, Pekka Ruusuvuori

**Institute of Biomedicine, University of Eastern Finland, Kuopio, Finland**
Sonja Koivukoski, Leena Latonen

**Department of Surgical Pathology, Zealand University Hospital, Roskilde, Denmark**
Dusan Rasic, Sandra Kristiane Sinius Pouplier, Anne-Vibeke Laenkholm

**Department of Oncology and Pathology, Karolinska Institutet, Stockholm, Sweden**
Johan Hartman

**Faculty of Medicine and Health Technology, Tampere University, Tampere, Finland**
Pekka Ruusuvuori

**Foundation for the Finnish Cancer Institute, Helsinki, Finland**
Leena Latonen

**MedTechLabs, BioClinicum, Karolinska University Hospital, Stockholm, Sweden**
Mattias Rantalainen, Johan Hartman

## Contributions

P.W., M.V., L.S. jointly organised the ACROBAT challenge and contributed equally to it. M.R., P.R. jointly supervised the ACROBAT challenge organisation. P.W., M.R., M.V. and P.R. jointly conceptualised the ACROBAT challenge. P.W. selected and verified the data set. P.W. drafted the manuscript. P.W., K.K., M.V. processed the images. L.S., M.V. implemented the annotation infrastructure. M.V., L.S., P.W. generated the annotation instructions. C.C., C.B., S.K., A.K., D.R., Y.F., S.P., P.W., M.V., L.S., K.K., A.S., K.L.E. (in order of contribution) generated the landmark annotations. M.R., J.H., P.R., A.L., L.L. acquired funding for this project. All authors contributed to editing the manuscript.

## Corresponding authors

Mattias Rantalainen (mattias.rantalainen@ki.se), Philippe Weitz (philippe.weitz@ki.se)

## Competing interests

M.R., J.H are co-founders and shareholders of Stratipath.